\documentclass[12pt]{article}
\pdfoutput=1
\usepackage{graphicx}
\usepackage{jheppub}
\usepackage{ytableau}
\usepackage{pgf,pgfarrows,pgfnodes}
\usepackage{graphics}
\usepackage{hyperref}
\usepackage{tikz}
\usepackage{xparse}
\usepackage{genyoungtabtikz}
\usepackage{subfigure}
\title{The light asymptotic limit of conformal
blocks in $\mathcal{N}=1$ super Liouville field theory}
\author{Hasmik Poghosyan}
\affiliation{Yerevan Physics Institute\\
Alikhanian Br. 2, 0036 Yerevan, Armenia}
\emailAdd{hasmikpoghos@gmail.com}
\abstract{ Analytic expressions for the two dimensional $\mathcal{N}=1$ SLFT blocks in the light semi-classical limit are found for  both Neveu-Schwarz and Ramond sectors. The  calculations are done by using the duality between $SU(2)$ $\mathcal{N}=2$ super-symmetric gauge theories living on $R^4/Z_2$ space and two dimensional  $\mathcal{N}=1$ super Liouville field theory. It is shown that in the light asymptotic limit only a restricted set of Young diagrams contribute to the partition function. This  enables us to sum up the instanton series explicitly and find closed expressions for the corresponding  $\mathcal{N}=1$ SLFT  four point blocks in the light asymptotic limit.}

\begin{document}
\maketitle
\newcommand{\ie}{{\it i.e.\ }}
\def\bea{\begin{eqnarray}}
\def\eea{\end{eqnarray}}
\def\a{\alpha}
\def\b{\beta}
\def\g{\gamma}
\def\G{\Gamma}
\def\d{\delta}
\def\D{\Delta}
\def\e{\epsilon}
\def\z{\zeta}
\def\th{\theta}
\def\k{\kappa}
\def\l{\lambda}
\def\m{\mu}
\def\n{\nu}
\def\r{\rho}
\def\s{\sigma}
\def\t{\tau}
\def\f{\phi}
\newpage
\section{Introduction}
Two dimensional conformal field theory (CFT) \cite{Belavin:1984vu} is relevant in statistical physics while studying second order phase transitions, and also it is an  important building block in String theory \cite{green1988superstring}.
 An important example of CFT is Liouville field theory (LFT) \cite{Polyakov:1981rd} which is a bosonic field   theory with exponential interaction. This theory is endowed with the spin two conserved currents that are the holomorphic and anti-holomorphic components of the stress energy tensor. The Fourier components of these currents obey the Virasoro algebra. There are more general CFTs which in addition to the spin two currents include also conserved currents with higher spins \cite{Zamolodchikov:1985wn}. The corresponding symmetry algebra is called $\mathcal{W}$ algebra. Important examples of theories that enjoy $\mathcal{W}$ symmetry  are Toda field theories. These theories  generalize  LFT to the case of several interacting scalar fields.

As a first step on the way of constructing a full fledged quantum theory it is instructive to investigate its quasi classical limit.
In both Liouville and Toda theories one can distinguish three types of quasi classical limits. These are mini-superspace, heavy and light limits. All three are  large central charge limits. They differ from each other by the behavior of primary fields under consideration. The primary fields are given by the vertex operators  $V_{\alpha}=e^{i\alpha \phi}$. In the light limit we choose $\alpha = \eta b$ and send $b$ to zero. Thus we take the large central charge limit keeping the conformal dimension finite.

The AGT  correspondence \cite{Alday:2009aq}  connects $2d$ conformal blocks in LFT to the Nekrasov Partition Function \cite{Nekrasov:2002qd,Losev:1997wp, Nekrasov:2003rj} of the four-dimensional $\mathcal{N}=2$ supersymmetric gauge theories. The AGT correspondence is a powerful tool not only for deriving correlation functions in $2d$ CFTs but also for studying gauge theories by applying CFT methods. The Nekrasov partition function can be represented as a sum over Young diagrams \cite{Nekrasov:2002qd, Flume:2002az, Nekrasov:2003rj} which   according to the AGT correspondence can be used to compute conformal blocks in $2d$ LFTs.
In \cite{Poghosyan:2016lya}  the $U(N)$ Nekrasov partition function in the light asymptotic limit was considered. It was proved that in this limit for a specific choice of fields in the Nekrasov partition function contribute only Young diagrams whose number of rows does not exceed $(N-1)$. This simplification makes it possible to write an explicit formula for the partition function in this limit.  After applying AGT duality a large class of $\mathcal{W}_N$ light conformal blocks for arbitrary $N$'s has been obtained.

$\mathcal{N} = 1$ super Liouville field theory  (SLFT) \cite{Polyakov:1981re}  is an important example of $\mathcal{N} = 1$ super conformal field theory (SCFT) \cite{Zamolodchikov:1988nm,Friedan:1984rv,Bershadsky:1985dq,Eichenherr:1985cx}. In \cite{Belavin:2012aa,Belavin:2011tb}
 an AGT like correspondence  between the $\mathcal{N} = 1$ SLFT and the  $U(2)$ super-symmetric gauge
theories living on the space $R^4/Z_2$  is given.

Besides the spin two conserved currents (energy-momentum tensor) SLFT  includes also spin $3/2$ currents (the super-currents). These currents generate super conformal symmetry which in $2d$ is described by the Neveu-Schwarz-Ramond algebra \cite{Zamolodchikov:1988nm, Eichenherr:1985cx, Bershadsky:1985dq}. If upon encircling a field by the super-current an extra multiplier $-1$ is produced, one refers to this field as a Ramond field. Those fields which are local with respect to the super current are called Neveu-Schwarz fields.\\
 In this paper different $\mathcal{N} = 1$ SLFT blocks in the light limit  are derived by using the above mentioned duality between super Yang-Mills theory and $2d$ SCFT.
We obtained that in the case of SLFT the analysis of the light limit is more subtle and complicated compare to the bosonic Lioville theory. In particular we found that 
in the light limit to the conformal blocks contribute  
 not only one row diagrams. For instance the instanton partition functions that correspond to the conformal blocks with four Ramond  fields also get contribution from  diagrams, like those in figures (\ref{figYDSL:subfigure2}) and (\ref{figYDSL:subfigure3}) below.

The paper is organized as follows.
In section \ref{Sintpartfunc}  the expression for the instanton partition functions of ${\cal N }=2$ SYM on $R^4/Z_2$ \cite{Fucito:2004ry, Fucito:2006kn}  is reviewed.
In section \ref{SLFT} we bring known facts for  ${\cal N }=1$ SLFT and its light asymptotic
limit that will be useful for us.
In subsection \ref{map}  the map between  ${\cal N }=1$ super Liouville conformal blocks and ${\cal N }=2$ SYM on $R^4/Z_2$ is given.
In subsection \ref{lalita} the rules for the light asymptotic limit are written.
In section \ref{LPF}  we present new results on  various partition function in the light limit.
In section \ref{CBSLFT}  by using these partition functions we give the corresponding  conformal blocks in the light limit.
In appendix \ref{ap0} some technical points on the instanton partition function of $SU(2)$ gauge  theories  on $R^4/Z_2$ are reviewed.
In appendix \ref{Appendix1} we proved that in the light limit to the instanton partition function  contribute only  the  Young diagrams depicted at figure \ref{figYD3} and, in appendix 	\ref{Appendix2}
computations of these partition functions in the light limit are given.
\section{The partition functions of $\mathcal{N}=2$ SYM on $R^4/Z_2$}
\label{Sintpartfunc}
Let us consider ${\cal N}=2 $ SYM theory  with a $U(2)$ gauge group on
the space $R^4/Z_2$.
The instanton part of the
partition function for this theory can be represented as (see \cite{Fucito:2004ry,Fucito:2006kn})
\begin{eqnarray}\label{instpartition}
Z_{(u_1,u_2),(v_1,v_2)}^{(q_1,q_2)}(\vec{a}^{(0)}, \vec{a}^{(1)}, \vec{a}^{(2)}|q)=
\sum_{\{\vec{Y}^{\vec{q}}\}}F_{\vec{Y}\,(u_1,u_2),(v_1,v_2)}^{\,\,\,\,\,(q_1,q_2)}\left(\vec{a}^{(0)}, \vec{a}^{(1)}, \vec{a}^{(2)}\right)
q^{\frac{|\vec{Y}|}{2}}\,.
\end{eqnarray}
The sum goes over the pairs of Young diagrams $\vec{Y}^{\vec{q}}=(Y^{q_1}_1,Y^{q_2}_2)$ colored in chess like order. To each diagram one ascribes a $\mathbb{Z}_2$
charge $q_i$, $i=1,2$ which indicates the color of the corner and takes values $0$ or  $1$ (white or black correspondingly).
$|\vec{Y}|$ is the total number of boxes in $Y_1$ and $Y_2$.
$q$ is the instanton counting parameter. Let us clarify our conventions on gauge theory parameters $a_i^{(0,1,2)}$,
$i=1,2 $. The parameters $a_i^{(1)}$ are expectation values
of the scalar field in vector multiplet. Without loss of generality
we will assume that the ``center of mass" of these expectation values is zero
\bea\label{au1}
{\bar a}^{(1)}=\frac{1}{2}\left( a_1^{(1)}+a_2^{(1)}\right)=0\,,
\eea
since a nonzero center of mass can
be absorbed by shifting hypermultiplet
masses. Furthermore $a_i^{(0)}$ ($a_i^{(2)}$) are the masses of fundamental
(anti-fundamental) hypers. \\
The expansion coefficient of the instanton partition function (\ref{instpartition}) is given by
\begin{eqnarray}
\label{F}
& F_{\vec{Y}\,(u_1,u_2),(v_1,v_2)}^{\,\,\,\,\,(q_1,q_2)}\left(\vec{a}^{(0)},
\vec{a}^{(1)}, \vec{a}^{(2)}\right)=\\
& \displaystyle\prod_{i=1}^2 \displaystyle\prod_{j=1}^2
\frac{Z_{bf}(u_i,a_i^{(0)},\varnothing\mid q_j, a_j^{(1)},Y_j)
Z_{bf}(q_i, a_i^{(1)},Y_i\mid v_j,a_j^{(2)},\varnothing)}
{Z_{bf}(q_i,a_i^{(1)},Y_i\mid q_j,  a_j^{(1)},Y_j)}\,,\nonumber
\end{eqnarray}
where
\newcount\tableauRow
\newcount\tableauCol
\def\tableauDim{0.4}
\newenvironment{Tableau}[1]{%
  \tikzpicture[scale=0.7,draw/.append style={loosely dotted,gray},
                      baseline=(current bounding box.center)]
    \tableauRow=-1.5
    \foreach \Row in {#1} {
       \tableauCol=0.5
       \foreach\k in \Row {
         \draw[thin](\the\tableauCol,\the\tableauRow)rectangle++(1,1);
         \draw[black,ultra thick](\the\tableauCol,\the\tableauRow)+(0.5,0.5)node{$\k$};
         \global\advance\tableauCol by 1
       }
       \global\advance\tableauRow by -1
    }
}{\endtikzpicture}
\newcommand\tableau[1]{\begin{Tableau}{#1}\end{Tableau}}
\begin{figure}
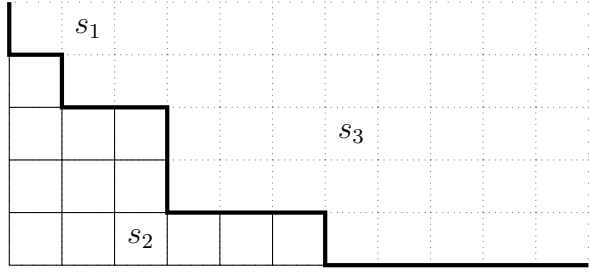

\center
\begin{tabular}{l@{\qquad}l@{\qquad}l}
\begin{Tableau}{{,s_1,,,,,,,,,},{,,,,,,,,,,},{,,,,,,s_3,,,,},
  {,,,,,,,,,,},{,,s_2,,,,,,,,}}
    \draw[ultra thick,solid ,color=black](11,-5)--(6,-5)
    --(6,-4)--(3,-4)--(3,-2)--(1,-2)--(1,-1)--(0,-1)--(0,0);
     \draw[ solid ,color=black](0,-5)--(0,-1);
     \draw[ solid ,color=black](1,-5)--(1,-1);
     \draw[ solid ,color=black](2,-5)--(2,-2);
     \draw[ ,solid ,color=black](3,-5)--(3,-4);
     \draw[ solid ,color=black](4,-5)--(4,-4);
     \draw[ solid ,color=black](5,-5)--(5,-4);
     \draw[solid ,color=black](0,-4)--(3,-4);
     \draw[ solid ,color=black](0,-3)--(3,-3);
     \draw[ solid ,color=black](0,-2)--(1,-2);
     \draw[solid ,color=black](0,-5)--(6,-5);
\end{Tableau}
\end{tabular}
\caption{Arm and leg length with respect to the Young diagram whose
 borders are
 outlined by dark black: $A(s_1)=-2$, $L(s_1)=-2$, $A(s_2)=2$, $L(s_2)=3$, $A(s_3)=-3$,
  $L(s_3)=-4$. }
\label{Fig:arm_and_leglenth}
\end{figure}
\begin{eqnarray}
\label{Zbf}
&Z_{bf}(x,a,\lambda\mid y,b,\mu)=\\
&\displaystyle\prod_{s\in\lambda^*}\big(a-b-\epsilon_1L_{\mu}(s)+
\epsilon_2(1+A_{\lambda}(s))\big)
\displaystyle\prod_{s\in\mu^*}\big(a-b+\epsilon_1(1+L_{\lambda}(s))
-\epsilon_2A_{\mu}(s)\big)\,.\nonumber
\end{eqnarray}
Here  $\epsilon_1$ and $\epsilon_2$ are the
$\Omega$-background parameters. We will use the notation
$\epsilon=\epsilon_1+\epsilon_2$.
$A_{\lambda}(s)$  ($L_{\lambda}(s)$) is  the arm-length (leg-length)
of the square $s$ towards the Young diagram $\lambda$, defined as oriented vertical
(horizontal) distance
of the square $s$ to outer boundary of the Young tableau $\lambda$
(see figure \ref{Fig:arm_and_leglenth}).
 $\lambda^*$,  $\mu^*$  are subsets of boxes $\lambda$ and $\mu$ respectively such that, 
a box of $\lambda$ ($\mu$) belongs to
 $\lambda^*$ ($\mu^*$) if and only if the  replacement
\bea
\epsilon_1,\, \epsilon_2\rightarrow 1;\,\, a\rightarrow x;\,\,
b\rightarrow y\,\,\,\,(i=1,2)
\eea
in the first (second) multiplier of (\ref{Zbf}) results in $0\, (mod \,2)$  (remind that $u_i$ and $v_i$ ($i=1,2$) take values $0$ or $1$). For more details see Appendix \ref{ap0}.

According to the duality between $\mathcal{N}=2$ SYM on $R^4/Z_2$ and $\mathcal{N}=1$ SLFT
these partition functions are directly related to
four point conformal blocks in ${\cal N }=1$ SLFT. Before
describing this relation let us briefly recall few facts about ${\cal N} =1$ SLFT itself.
\section{Known  facts on ${\cal N }=1$ SLFT and its light asymptotic
limit}
\label{SLFT}
Super-Liouville field theory is a supersymmetric generalization of the bosonic Liouville theory, which is known to be the theory of matter induced gravity in two dimensions.
Similarly SLFT describes 2d supergravity, induced by supersymmetric matter.
Super-Liouville field theory on a two-dimensional surface with metric $g_{ab}$ is given by the Lagrangian density
\bea
{\cal L}={1\over 2\pi}g^{ab}\partial_a\varphi\partial_b \varphi+
{1\over 2\pi}(\psi\bar{\partial}\psi+\bar{\psi}\partial\bar{\psi})+2i\mu b^2\bar{\psi}\psi e^{b\varphi}+
2\pi \mu^2 b^2 e^{2b\varphi}\, .
\eea
There are two kinds of fields in $2d$ ${\cal N } =1$ SLFT called  Neveu-Schwarz  and Ramond fields to be specified below.
The symmetries of the  theory are generated by the
energy-momentum tensor and the superconformal currents
\bea
&&T=-{1\over 2} (\partial\varphi\partial\varphi-Q\partial^2\varphi+\psi\partial\psi)\,,\\
&&G=i(\psi\partial\varphi-Q\partial\psi)\, .
\eea
Commutation relation of the Neveu-Schwarz-Ramond algebra are
 \bea
 \label{VIRALG}
&& [L_m,L_n]=(m-n)L_{m+n}+{c\over 12}m(m^2-1)\delta_{m+n}\, ,\\
\label{LGRALG}
 &&[L_m, G_k]={m-2k\over 2}G_{m+k}\,,\\
 \label{SUSYALG}
 &&\{G_k,G_l\}=2L_{l+k}+{c\over 3}\left(k^2-{1\over 4}\right)\delta_{k+l}\, ,
 \eea
with the central charge
\bea
c_L={3\over 2}+3Q^2\, ,
\text{ where }
Q=b+b^{-1}\, .
\eea
Here $L_m$ and $G_k$ are the Laurent series coefficients of the currents $T$ and $G$ respectively.  
For the Ramond algebra $k$ and $l$ take integer and for the Neveu-Schwarz algebra half-integer values.

It is known that in the Neveu-Schwarz sector at the light asymptotic limit  the symmetry algebra reduces to the finite subalgebra generated by  $L_0\,,L_{\pm1}\,,G_{\pm1/2}$ only.
Notice that for this subalgebra  the central extension terms in (\ref{VIRALG}) and (\ref{SUSYALG}) disappear.
  For the mentioned values
of $m$ and $l$ (\ref{VIRALG})-(\ref{SUSYALG}) is obviously closed. For the Ramond sector its light asymptotic limit is more
subtle and needs to be clarified yet.

NS primary fields $\Phi_{\alpha}(z,\bar{z})$ in this theory,
$\Phi_{\alpha}(z,\bar{z})=e^{\alpha \varphi(z,\bar{z})}$, have conformal dimensions
\bea
\label{NSdim}
\Delta^{NS}_{\alpha}={1\over 2}\alpha(Q-\alpha)\, .
\eea
Introduce also the field that is the highest component of the NS superfield  build from $\Phi_{\alpha}$
\bea
\Phi_{\tilde{\alpha}}(z,\bar{z})=G_{-1/2}\bar{G}_{-1/2}\Phi_{\alpha}(z,\bar{z})\,,
\eea
with dimension 
\bea
\label{NSTildedim}
\tilde{\Delta}^{NS}_{\alpha}=\Delta^{NS}_{\alpha}+1/2\, ,
\eea
and as well as the Ramond primary fields defined as 
\bea
R_{\alpha}^{\pm}(z,\bar{z})=\sigma^{\pm}(z,\bar{z}) e^{\alpha \varphi(z,\bar{z})}
\eea
where $\sigma^{\pm}$ is the spin field with dimension $1/16$. 
Thus the dimension of a Ramond  operator is
\bea
\label{Rdim}
\Delta_{\alpha}^R={1\over 16}+{1\over 2}\alpha(Q-\alpha)\, .
\eea
\\
\section{${\cal N }=1$ Super Liouville conformal blocks and their relation to the ${\cal N }=2$ SYM on $R^4/Z_2$}
\label{agtcft}
Let us schematically denote by $\langle \Psi_1(\infty)\Psi_2(1)\Psi_3(q)\Psi_4(0)\rangle_{\Delta^{\Psi}}$
conformal block of $\Psi_i$, $i=1\ldots 4$, fields with  intermediary field $\Psi$ of conformal weight 
$\Delta^{\Psi}$.

Four point blocks where  all four fields are bosonic primaries $\Phi_i$ with conformal
 weights $\Delta_{\alpha_i}$   are connected with the
$Z_{inst}$ partition function in the following  way (see \cite{Belavin:2011tb})
\bea
\label{AGTbos1}
^{\lozenge}Z_{(0,0),(0,0)}^{(0,0)}=q^{\Delta^{NS}_{1}+\Delta^{NS}_{2}-\Delta^{NS}}(1-q)^U
\langle \Phi_4(\infty)\Phi_3(1)
\Phi_1(q)
\Phi_2(0)\rangle_{\Delta^{NS}}
\eea
and for $\tilde{\Delta}=\Delta+\frac{1}{2}$
\bea
\label{AGTbos2}
^{\blacklozenge}Z_{(0,0),(0,0)}^{(1,1)}=\frac{q^{\Delta^{NS}_{1}+\Delta^{NS}_{2}-\tilde{\Delta}^{NS}}}{2}(1-q)^U
\langle \Phi_4(\infty)\Phi_3(1)\Phi_1(q)\Phi_2 (0)\rangle_{\tilde{\Delta}^{NS}}\,.
\eea
The index $^{\lozenge} $ shows that the number of black and white boxes (the number of boxes in
 both diagrams together) are equal  and the index $^{\blacklozenge}$ show the number differ by one.
In the expressions (\ref{AGTbos1}) and (\ref{AGTbos2}) $U$ is given by
\bea
U=\alpha_2\left(Q-\alpha_3\right)\,.
\label{U}
\eea
We will see that in the light asymptotic limit $U$  is just one. So in this limit the
 corresponding partiton function gives the four point conformal block for bosonic
 fields.\\
Let us look at the $\langle R \Phi \Phi R \rangle$ type conformal block. According to
 \cite{Belavin:2012aa} this conformal blocks are connected to the instanton
partition function in the following way
\bea
\label{AGTbosfer++}
&&^{\lozenge}Z_{(0,0),(0,0)}^{(0,1)}=q^{\Delta^{R}_{3}+\Delta^{NS}_{4}-\Delta^{R}}(1-q)^{(U-\frac{3}{8}+d_1-d_2-d_3+d_4)}
\langle R^{+}_2(\infty) \Phi_1(1) \Phi_4(q) R^{+}_3(0) \rangle _{\Delta^R}\,.\,\,\,\quad
\eea
Now let us look at the $\langle R R R R \rangle$ conformal blocks \cite{Belavin:2012aa}.
For the partition functions with equal numbers of black and white cells
\bea
\label{Zfor_equal}
&&^{\lozenge}Z_{(1,0),(1,0)}^{(0,0)}(q)=(1-q)^U
\left(G_{sl(2)}(q)H_{-}(q)+\tilde{G}_{sl(2)}(q)\tilde{H}_{-}(q)\right)\,,\\
&&^{\lozenge}Z_{(0,1),(0,1)}^{(0,0)}(q)=(1-q)^U
\left(G_{sl(2)}(q)H_{+}(q)+\tilde{G}_{sl(2)}(q)\tilde{H}_{+}(q)\right)\,,\\
&&^{\lozenge}Z_{(1,0),(0,1)}^{(0,0)}(q)=(1-q)^U
\left(G_{sl(2)}(q)F_{-}(q)+\tilde{G}_{sl(2)}(q)\tilde{F}_{-}(-q)\right)\,,\\
&&^{\lozenge}Z_{(0,1),(1,0)}^{(0,0)}(q)=(1-q)^U
\left(G_{sl(2)}(q)F_{+}(q)+\tilde{G}_{sl(2)}(q)\tilde{F}_{+}(-q)\right)\,.
\eea
For the partition functions whose numbers of  black and white boxes differ by one
\bea
&&^{\blacklozenge}Z_{(1,0),(1,0)}^{(1,1)}(q)=(1-q)^U
\left(\tilde{G}_{sl(2)}(q)H_{+}(q)+G_{sl(2)}(q)\tilde{H}_{+}(q)\right)\,,\\
&&^{\blacklozenge}Z_{(0,1),(0,1)}^{(1,1)}(q)=(1-q)^U
\left(\tilde{G}_{sl(2)}(q)H_{-}(q)+G_{sl(2)}(q)\tilde{H}_{-}(q)\right)\,,\\
&&^{\blacklozenge}Z_{(1,0),(0,1)}^{(1,1)}(q)=(1-q)^U
\left(\tilde{G}_{sl(2)}(q)F_{+}(q)+G_{sl(2)}(q)\tilde{F}_{+}(-q)\right)\,,\\
&&^{\blacklozenge}Z_{(0,1),(1,0)}^{(1,1)}(q)=(1-q)^U
\left(\tilde{G}_{sl(2)}(q)F_{-}(q)+G_{sl(2)}(q)\tilde{F}_{-}(-q)\right)\,.
\label{Zfor_diff1}
\eea
Here $H_{\pm}$, $F_{\pm}$, $\tilde{H}_{\pm}$ and $\tilde{F}_{\pm}$ are related to the conformal blocks
 containing  four  Ramond fields, for their definition see \cite{Belavin:2012aa}. $G(q)$ and  $\tilde{G}(q)$ are given by
\bea
&G(q)=(1-q)^{-\frac{3}{8}} \sqrt{\frac{1}{2} \left(1+\sqrt{1-q}\right)}\,, \\
 &\tilde{G}(q)=(1-q)^{-\frac{3}{8}} \sqrt{\frac{1}{2} \left(1-\sqrt{1-q}\right)}\,.
\eea
\\
Below is given the map that  connects the instanton partition functions of ${\cal N }=2$ SYM on $R^4/Z_2$ to the ${\cal N }=1$ SLFT conformal blocks.
\subsection{The map relating partition functions to conformal blocks}
\label{map}
First of all,
the instanton counting parameter $q$ gets identified with the cross
ratio of insertion points, as already anticipated in formulas (\ref{Zfor_equal})-(\ref{Zfor_diff1}),  in CFT block. The Liouville parameter $b$ is related to
the $\Omega$-background parameters via
\bea
b=\sqrt{\frac{\epsilon_1}{\epsilon_2}}\,.
\label{bepsilon}
\eea
\begin{figure}[t]
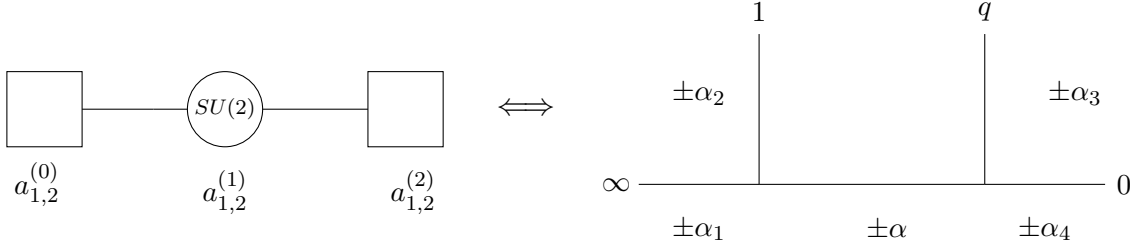

\begin{pgfpicture}{0cm}{0cm}{15cm}{4cm}

\pgfcircle[stroke]{\pgfpoint{3cm}{2.8cm}}{0.5cm}
\pgfputat{\pgfxy(3,2.8)}{\pgfbox[center,center]{\scriptsize{$SU(2)$}}}
\pgfputat{\pgfxy(3,1.7)}{\pgfbox[center,center]{\small{$a_{1,2}^{(1)}$}}}
{\color{black}\pgfrect[stroke]{\pgfpoint{0.1cm}{2.3 cm}}{\pgfpoint{1cm}{1cm}}}
\pgfputat{\pgfxy(0.5,1.8)}{\pgfbox[center,center]{\small{$a_{1,2}^{(0)}$}}}
{\color{black}\pgfrect[stroke]{\pgfpoint{4.9cm}{2.3cm}}{\pgfpoint{1cm}{1cm}}}
\pgfputat{\pgfxy(5.5,1.7)}{\pgfbox[center,center]{\small{$a_{1,2}^{(2)}$}}}
\pgfline{\pgfxy(1.1,2.8)}{\pgfxy(2.05,2.8)}
\pgfline{\pgfxy(2.05,2.8)}{\pgfxy(2.5,2.8)}
\pgfline{\pgfxy(3.5,2.8)}{\pgfxy(4.9,2.8)}
\pgfputat{\pgfxy(7,2.8)}{\pgfbox[center,center]{$\Longleftrightarrow$}}
\pgfline{\pgfxy(8.5,1.8)}{\pgfxy(10.1,1.8)}
\pgfline{\pgfxy(10.1,1.8)}{\pgfxy(10.1,3.8)}
\pgfline{\pgfxy(10.1,1.8)}{\pgfxy(13.1,1.8)}
\pgfline{\pgfxy(13.1,1.8)}{\pgfxy(13.1,3.8)}
\pgfline{\pgfxy(13.1,1.8)}{\pgfxy(14.7,1.8)}
\pgfputat{\pgfxy(11.8,1.2)}{\pgfbox[center,center]{\small{$\pm\alpha $}}}
\pgfputat{\pgfxy(9.3,3)}{\pgfbox[center,center]{\small{$\pm\alpha_2 $}}}
\pgfputat{\pgfxy(9.3,1.2)}{\pgfbox[center,center]{\small{$\pm\alpha_1 $}}}
\pgfputat{\pgfxy(14.3,3)}{\pgfbox[center,center]{\small{$\pm\alpha_3 $}}}
\pgfputat{\pgfxy(13.9,1.2)}{\pgfbox[center,center]{\small{$\pm\alpha_4 $}}}
\pgfputat{\pgfxy(8.2,1.8)}{\pgfbox[center,center]{\small{$\infty$}}}
\pgfputat{\pgfxy(10.1,4.1)}{\pgfbox[center,center]{\small{$1$}}}
\pgfputat{\pgfxy(13.1,4.1)}{\pgfbox[center,center]{\small{$q$}}}
\pgfputat{\pgfxy(14.95,1.8)}{\pgfbox[center,center]{\small{$0$}}}
\pgfclearendarrow
\end{pgfpicture}
\caption{On the left: the quiver diagram for the conformal $SU(2)$ gauge theory.
On the right: the diagram of the conformal block for the dual ${\cal N }=1$ SLFT .}
\label{figAGT}
\end{figure}
The map between the gauge parameters (\ref{instpartition})  and conformal block parameters
can be established from the following rules (see Fig.\ref{figAGT}).
 First define the rescaled gauge parameters
\bea\label{renga}
A_i^{(0)}=\frac{a_i^{(0)}}{\sqrt{\epsilon_1\epsilon_2}}\,;
\quad\quad A_i^{(1)}=\frac{a_i^{(1)}}{\sqrt{\epsilon_1\epsilon_2}}\,;\quad
\quad A_i^{(2)}=\frac{a_i^{(2)}}{\sqrt{\epsilon_1\epsilon_2}}\,,
\eea    
where $i=1,2$.

	Then
\begin{itemize}            
\item
The differences between the ``centers of masses" of the successive rescaled gauge
 parameters (\ref{renga}) give the charges of  the ``vertical" entries of the
  conformal block:
\bea
\bar{A}^{(1)}-\bar{A}^{(0)}=\alpha_2\, ;\quad\quad \bar{A}^{(2)}-\bar{A}^{(1)}=
\alpha_3\,.
\eea
\item
The rescaled gauge parameters with the subtracted centers of masses give the momenta of
 the ``horizontal" entries of the conformal block:
\bea\label{hormom}
&&A_i^{(0)}-\bar{A}^{(0)}=(-)^{i+1}\left(\alpha_1-\frac{Q}{2}\right)\, ;\\ \nonumber
 &&A_i^{(1)}-\bar{A}^{(1)}=(-)^{i+1}\left(\alpha-\frac{Q}{2}\right)\,; \\ \nonumber
&&A_i^{(2)}-\bar{A}^{(2)}=(-)^{i+1}\left(\alpha_4-\frac{Q}{2}\right)\, .
\eea
\end{itemize}
Using (\ref{au1}) and (\ref{renga})-(\ref{hormom}) we obtain the relation
between the gauge and conformal parameters:
\begin{eqnarray}
\label{au}
\frac{a_i^{(0)}}{\sqrt{\epsilon_1\epsilon_2}}&=&(-)^{i+1}\left(\alpha_1
-\frac{Q}{2}\right)-\alpha_2\,;\nonumber\\
\frac{a_i^{(1)}}{\sqrt{\epsilon_1\epsilon_2}}&=&(-)^{i+1}\left(\alpha
-\frac{Q}{2}\right)\,;\\
\frac{a_i^{(2)}}{\sqrt{\epsilon_1\epsilon_2}}&=&(-)^{i+1}\left(\alpha_4
-\frac{Q}{2}\right)+\alpha_3\,.\nonumber
\end{eqnarray}
\subsection{Light asymptotic limit of the gauge parameters }
\label{lalita}
In this paper we are interested in so called "light" asymptotic
limit i.e. the central charge is sent to infinity (i.e. $b\rightarrow 0$)
while keeping the dimensions finite. It follows from (\ref{NSdim}) and (\ref{Rdim})
that
to reach this limit one can simply put
\bea
\alpha=b \eta;\qquad \alpha_l=b \eta_l; \quad  \text{where} \quad l=1; \, 2;\;4,
\label{eta14}
\eea
by keeping all the parameters $\eta$ finite.
If we exchange  $\alpha$ with $Q-\alpha$ the conformal dimension remains the same
 (see (\ref{NSdim}) and (\ref{Rdim})), so for $\alpha_3$ we can take as its
 light asymptotic limit
\bea
Q-\alpha_3=b \eta_3
\label{eta2}
\eea
By taking the limit in this way we get rid of the $U(1)$ factor defined in (\ref{U}).
Using (\ref{eta14}), (\ref{eta2})
 we can rewrite the AGT map (\ref{au}) as
\begin{eqnarray}
\label{au_light0}
a_i^{(0)}&=&(-)^{i+1}\left(\epsilon_1\eta_1-\frac{\epsilon}{2}
\right)-\epsilon_1\eta_2 \, \\
\label{au_light1}
a_i^{(1)}&=&(-)^{i+1}\left(\epsilon_1\eta-\frac{\epsilon}{2}\right) \, ;\\
\label{au_light2}
a_i^{(2)}&=&(-)^{i+1}\left(\epsilon_1\eta_4-\frac{\epsilon}{2}
\right)+\epsilon -\epsilon_1 \eta_3\, .
\end{eqnarray}
\section{Partition function in the light asymptotic limit}
\label{LPF}
\begin{figure}
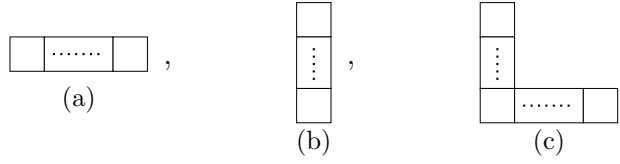

\centering
\Yvcentermath1
\newcommand\sesqui{1.5}
\newcommand\eleven{11}
\subfigure[]
{\gyoung(<>_2\hdts<>)
\label{figYDSL:subfigure1}}\,,\qquad\qquad
\subfigure[]
{\gyoung(<>,|\sesqui\vdts,<>)
\label{figYDSL:subfigure2}}\,,\qquad\qquad
\subfigure[]
{\gyoung(<>,|\sesqui\vdts,<>_2\hdts<>)
\label{figYDSL:subfigure3}}
\caption{the possible nonempty Young diagrams}\label{figYD3}
\end{figure}
We have shown in appendix \ref{Appendix1} that for
the light asymptotic limit only a restricted set of Young diagrams contribute to the instanton partition function.
This set varies  depending on the charges  and  the differences of black and white cells of the related Young diagrams.
Below are given all
pairs of $Y_1$ and $Y_2$ for which the coefficient of the instanton expansion (\ref{instpartition}) is non zero in the light limit. In order to compute these coefficients for a
given pair of diagrams $Y_1$ and $Y_2$ one  makes use of (\ref{F}), (\ref{Zbf}), (\ref{au_light0})-(\ref{au_light2})
and then goes to the  light  limit $\epsilon_1\to 0$.
The results are given below (detailed calculation for some of the coefficients  can be found in appendix \ref{Appendix2}).
\subsection{Partition functions corresponding  to conformal blocks with four Neveu-Schwarz fields.
}
 The expansion coefficient $^{\lozenge}F^{(0,0)}_{(0,0),(0,0)}$ does not vanish in the light asymptotic limit if $Y_2$ is a empty Young diagram and $Y_1$ (see figure \ref{figYDSL:subfigure1})
has only one row  with $2k$ boxes, where $k$ can be zero or any positive integer.
It is equal to
\bea
\label{F(0,0)(0,0)(0,0)}
^{\lozenge}_LF^{(0,0)}_{(0,0),(0,0)}=\frac{
\left(\frac{1}{2}\left(\eta-\eta_4+\eta_3\right)\right)_k
\left(\frac{1}{2}\left(\eta-\eta_1+\eta_2\right)\right)_k
}{k! \left(\eta\right)_k}\,.
\eea
For more  details see appendix \ref{Appendix2}.\\
Inserting  (\ref{F(0,0)(0,0)(0,0)}) in  (\ref{instpartition}), we derive
\bea
\label{Z(0,0)(0,0)(0,0)}
^{\lozenge}_LZ^{(0,0)}_{(0,0),(0,0)}(q)=
\, _2F_1\left(A,B;\eta;q\right)\,.
\eea
Here  $A$ and $B$ are
\bea
A=\frac{1}{2}\left(\eta-\eta_1+\eta_2\right) \text{ and }
B=\frac{1}{2}\left(\eta-\eta_4+\eta_3\right)\,,
\eea
and $\, _2F_1(a,b;c;x)$ is the hypergeometric function.
It has the series expansion
\bea
\, _2F_1(a,b;c;x)=
\sum _{k=0}^{\infty } \frac{ (a)_k (b)_k}{k! (c )_k}x^k\,, \text{  where  } (u)_k=u(u+1)\dots(u+k-1)\,.
\eea
In the case of $^{\blacklozenge}F^{(1,1)}_{(0,0),(0,0)}$  for some set of pairs $Y_1$, $Y_2$ one gets large coefficients of order $\frac{1}{\e_1}$. Thus one should take into account these pairs and neglect those pairs whose
 contributions are of order $O(1)$ or bigger. An analyses quite similar to the one
 presented in  the appendix \ref{Appendix1}, shows that
$Y_2$ should be an empty and $Y_1$ must have
 a single row with $2k+1$ boxes  (see figure \ref{figYDSL:subfigure1}).Their contribution is
\bea
\label{(1,1)(0,0),(0,0)}
^{\blacklozenge}_LF^{(1,1)}_{(0,0),(0,0)}=\frac{1}{\e_1 \e_2}\frac{
\left(\frac{1}{2}\left(\eta-\eta_4+\eta_3+1\right)\right)_k
\left(\frac{1}{2}\left(\eta-\eta_1+\eta_2+1\right)\right)_k
}{2\, k! \left(\eta\right)_{k+1}}\,.
\eea
After inserting it in (\ref{instpartition}), we will get
\bea
\label{Z(1,1)(0,0),(0,0)}
^{\blacklozenge}_LZ^{(1,1)}_{(0,0),(0,0)}(q)=\frac{1}{\e_1 \e_2}
\frac{\sqrt{q}}{2 \eta}\,_2F_1\left(A+\frac{1}{2},B+\frac{1}{2};\eta+1;q\right)\,.
\eea

\subsection{Partition function  corresponding  to the conformal block with two Neveu-Schwarz and two Ramond  fields.}
The  coefficients of both  $^{\lozenge}Z^{(0,1)}_{(0,0),(0,0)}$ and $^{\lozenge}Z^{(1,0)}_{(0,0),(0,0)}$  do not vanish in the light limit if
 $Y_2$ is empty and $Y_1$ (see figure
 \ref{figYDSL:subfigure1}) is a diagram with only one row  with $2k$ boxes. Their contributions are
\bea
\label{F(1,0)(0,0)(0,0)}
^{\lozenge}_LF^{(1,0)}_{(0,0),(0,0)}=\frac{
\left(\frac{1}{2}\left(\eta-\eta_4+\eta_3+1\right)\right)_k
\left(\frac{1}{2}\left(\eta-\eta_1+\eta_2+1\right)\right)_k
}{ k! \left(\eta+\frac{1}{2}\right)_k}\, .
\eea
\bea
\label{F(0,1)(0,0)(0,0)}
^{\lozenge}_LF^{(0,1)}_{(0,0),(0,0)}=\frac{
	\left(\frac{1}{2}\left(\eta-\eta^{(4)}+\eta^{(3)}\right)\right)_k
	\left(\frac{1}{2}\left(\eta-\eta^{(1)}+\eta^{(2)}\right)\right)_k
}{ k! \left(\eta+\frac{1}{2}\right)_k}\,.
\eea
The corresponding  partition functions are
 \bea
 \label{Z(1,0)(0,0)(0,0)}
^{\lozenge}_LZ^{(1,0)}_{(0,0),(0,0)}(q)=
\,_2F_1\left(A+\frac{1}{2},
B+\frac{1}{2};
\eta+\frac{1}{2};q\right)\,.
 \eea
 \bea
 \label{Z(0,1)(0,0)(0,0)}
 ^{\lozenge}_LZ^{(0,1)}_{(0,0),(0,0)}(q)=
 \, _2F_1\left(A,B;\eta+\frac{1}{2};q\right)
 \eea
\subsection{Partition functions corresponding to conformal blokes with four Ramond fields.
}
 $^{\lozenge}F^{(0,0)}_{(0,1),(0,1)}$ differs from zero in the light asymptotic limit
if $Y_2$ (see figure \ref{figYDSL:subfigure2}) is a single column diagram with $2m$ boxes,
and  $Y_1$ (see figure \ref{figYDSL:subfigure1}) a single row diagram with
$2k$ boxes, where $m$ and $k$ can be zero or any positive integer. Their contribution is
\bea
\label{F(0,0)(0,1)(0,1)}
^{\lozenge}_LF^{(0,0)}_{(0,1),(0,1)}=
\left(\frac{\left(1/2\right)_m}{m!}\right)^2
\frac{\left(\frac{1}{2}\left(\eta-\eta_4+\eta_3\right)\right)_k
\left(\frac{1}{2}\left(\eta-\eta_1+\eta_2\right)\right)_k
}{ k! \left(\eta\right)_k}\,.
\eea
Its instanton partition function is
 \bea
 \label{Z(0,0)(0,1)(0,1)}
^{\lozenge}_LZ^{(0,0)}_{(0,1),(0,1)}(q)=
\frac{2}{\pi }K(q) \, _2F_1\left(A,B;\eta;q\right)\,.
 \eea
$K(x)$ and $E(x)$ are complete elliptic integrals of the first and second  kind correspondingly. They can be expressed in
terms of the Gauss hypergeometric function, as
\bea
K(x)=\frac{\pi}{2} \, _2F_1(\frac{1}{2},\frac{1}{2};1;x)\text{  and  }
E(x)=\frac{\pi}{2}   \, _2F_1\left(\frac{1}{2},-\frac{1}{2};1;x\right)
\eea
In the case of $^{\lozenge}F^{(0,0)}_{(1,0),(1,0)}$ for  pairs of Young diagrams $Y_2$, $Y_1$,
with $Y_2$  empty and
$Y_1$ (see figure \ref{figYDSL:subfigure3})   possessing  one column
with $2m$ boxes and other  $2k$ columns  with only one box,
 one gets large coefficients of order $\frac{1}{\e_1}$ in the light limit.
 In total $Y_1$  consists
 of $2m+2k$ boxes. These pairs give the main contribution. These terms are
\bea
\label{F(0,0)_(1,0),(1,0)}
^{\lozenge}_LF^{(0,0)}_{(1,0),(1,0)}=\frac{\epsilon_2}{\epsilon_1}
\frac{\left(\frac{1}{2}\right)_m\left(-\frac{1}{2}\right)_m}
{ (m-1)!m!}
\frac{\left(\frac{1}{2}\left(\eta-\eta_4+\eta_3+1\right)\right)_k
\left(\frac{1}{2}\left(\eta-\eta_1+\eta_2
+1\right)\right)_k
}{ k!\eta \left(\eta+1\right)_k}\,.
\eea
Its  partition function is given by
 \bea
 \label{Z(0,0)_(1,0),(1,0)}
^{\lozenge}_LZ^{(0,0)}_{(1,0),(1,0)}(q)=\frac{\epsilon_2}{\epsilon_1}
\frac{\left(E(q)-K(q)\right)}{\pi \eta}
\, _2F_1\left(A+\frac{1}{2},B+\frac{1}{2};\eta+1;q\right)\,.
 \eea
 
$^{\lozenge}_LF^{(0,0)}_{(0,1),(1,0)}$  differs from zero if $Y_2$ is empty and $Y_1$ is
a one row diagram (see figure \ref{figYDSL:subfigure1}) with $2k$ boxes. Their contribution is
\bea
\label{F(0,0)_(0,1),(1,0)}
^{\lozenge}_LF^{(0,0)}_{(0,1),(1,0)}=
\frac{\left(\frac{1}{2}\left(\eta-\eta_4+\eta_3+1\right)\right)_k
\left(\frac{1}{2}\left(\eta-\eta_1+\eta_2\right)\right)_k
}{ k! \left(\eta\right)_k}\,.
\eea
Its instanton partition function is given by
 \bea
\label{Z(0,0)_(0,1),(1,0)}
^{\lozenge}_LZ^{(0,0)}_{(0,1),(1,0)}(q)=
\, _2F_1\left(A,B+\frac{1}{2};\eta;q\right)\,.
 \eea
 
  $^{\lozenge}_LF^{(0,0)}_{(1,0),(0,1)}$ is not zero if $Y_2$ is empty and $Y_1$ (see figure \ref{figYDSL:subfigure1}) is
a one row diagram  with $2k$ boxes. Their contribution is
\bea
\label{F(0,0)_(1,0),(0,1)}
^{\lozenge}_LF^{(0,0)}_{(1,0),(0,1)}=
\frac{\left(\frac{1}{2}\left(\eta-\eta_1+\eta_2+1\right)\right)_k
\left(\frac{1}{2}\left(\eta-\eta_4+\eta_3\right)\right)_k
}{ k! \left(\eta\right)_k}\,.
\eea
Its partition function is given by
 \bea
 \label{Z(0,0)_(1,0),(0,1)}
^{\lozenge}_LZ^{(0,0)}_{(1,0),(0,1)}(q)=
\, _2F_1\left(A+\frac{1}{2},B;\eta;q\right)\,.
 \eea
In the case of $^{\blacklozenge}F^{(1,1)}_{(0,1),(0,1)}$ 
 for some set of pairs $Y_1$, $Y_2$ one gets large coefficients of order $\frac{1}{\e_1}$ in the light limit.
 These coefficients will give the main contribution in the partition function. These terms are obtained when $Y_2$ is empty and $Y_1$ (see figure \ref{figYDSL:subfigure3})
 has one column with $2m+1$ boxes and $2k$ columns with only one box, the
 total number of boxes is equal to $2m+2k+1$. They are given by
 \bea
 \label{F(1,1)_(0,1),(0,1)}
^{\blacklozenge}_LF^{(1,1)}_{(0,1),(0,1)}=\frac{\epsilon_2}{\epsilon_1}
\left(\frac{\left(\frac{1}{2}\right)_m}
{m!}\right)^2
\frac{\left(\frac{1}{2}\left(\eta-\eta_4+\eta_3+1\right)\right)_k
\left(\frac{1}{2}\left(\eta-\eta_1+\eta_2+1\right)\right)_k
}{-2\eta k! \left(\eta+1\right)_k}\,.
\eea
For its partition function, we receive
 \bea
 \label{Z(1,1)_(0,1),(0,1)}
^{\blacklozenge}_LZ^{(1,1)}_{(0,1),(0,1)}(q)=-\frac{\epsilon_2}{\epsilon_1}
\frac{\sqrt{q}}{\pi  \eta }K(q)
 \, _2F_1\left(A+\frac{1}{2},B+\frac{1}{2};\eta+1;q\right)\,.
 \eea
 
 $^{\blacklozenge}_LF^{(1,1)}_{(1,0),(1,0)}$  differs from zero if $Y_2$  is a one column
 diagram (see figure \ref{figYDSL:subfigure2}) with $2m+1$ boxes and $Y_1$ is a one
  row diagram (see figure \ref{figYDSL:subfigure1}) with $2k$ boxes. Their contribution is
\bea
\label{F(1,1)(1,0),(1,0)}
^{\blacklozenge}_LF^{(1,1)}_{(1,0),(1,0)}=
\frac{1}{(2+2m)(1+2m)}
\left(\frac{\left(\frac{3}{2}\right)_m}
{m!}\right)^2
\frac{\left(\frac{1}{2}\left(\eta-\eta_4+\eta_3\right)\right)_k
\left(\frac{1}{2}\left(\eta-\eta_1+\eta_2\right)\right)_k
}{ k! \left(\eta\right)_k}\,.
\nonumber\\
\eea
For the corresponding instanton partition function, we will get
\bea
\label{Z(1,1)(1,0),(1,0)}
^{\blacklozenge}_LZ^{(1,1)}_{(1,0),(1,0)}(q)=
-\frac{2 (E(q)-K(q))}{\pi  \sqrt{q}}
\, _2F_1\left(A,B;\eta;q\right)\,.
\eea

  Both $^{\blacklozenge}_LF^{(1,1)}_{(1,0),(0,1)}$ and
  $^{\blacklozenge}_LF^{(1,1)}_{(0,1),(1,0)}$   do not  vanish if
$Y_2$ is empty and $Y_1$ (see figure \ref{figYDSL:subfigure1})
is a one row diagram with $2k+1$ boxes. Their contributions are
\bea
\label{F(1,1)(0,1)(1,0)}
^{\blacklozenge}_LF^{(1,1)}_{(0,1),(1,0)}=
\frac{\left(\frac{1}{2}\left(\eta-\eta_1+\eta_2+1\right)\right)_k
\left(\frac{1}{2}\left(\eta-\eta_4+\eta_3\right)\right)_{k+1}
}{ k! \left(\eta\right)_k}\,,
\\
\label{F(1,1)(1,0)(0,1)}
^{\blacklozenge}_LF^{(1,1)}_{(1,0),(0,1)}=
\frac{\left(\frac{1}{2}\left(\eta-\eta_1+\eta_2\right)\right)_{k+1}
\left(\frac{1}{2}\left(\eta-\eta_4+\eta_3+1\right)\right)_{k}
}{ k! \left(\eta\right)_k}\,.
\eea
Their partition functions are
 \bea
 \label{Z(1,1)(0,1)(1,0)}
^{\blacklozenge}_LZ^{(1,1)}_{(0,1),(1,0)}(q)=
\frac{B}{\eta}\sqrt{q}\,
 \, _2F_1\left(A+\frac{1}{2},B+1;\eta+1;q\right)\,.
 \\
\label{Z(1,1)(1,0)(0,1)}
^{\blacklozenge}_LZ^{(1,1)}_{(1,0),(0,1)}(q)=
\frac{A}{\eta}\sqrt{q}\,
 \, _2F_1\left(A+1,B+\frac{1}{2};\eta+1;q\right)\,.
\eea
\section{Conformal blocks for ${\cal N }=1$ SLFT  in the light asymptotic limit}
\label{CBSLFT}
Applying (\ref{Z(0,0)(0,0)(0,0)}) and (\ref{Z(1,1)(0,0),(0,0)}) to (\ref{AGTbos1}) and (\ref{AGTbos2})
we will get the conformal blocks with all four fields being $NS$ in the light limit.
\bea
\label{NSNSNSNS(D)}
\langle \Phi_4(\infty)\Phi_3(1)\Phi_1(q)\Phi_2(0)\rangle^L_{\Delta^{NS}}
=q^{\frac{1}{2}(\eta-\eta^{(2)}-\eta^{(1)})}
\,_2F_1\left(A,B;\eta;q\right)
\eea
\bea
\label{NSNSNSNS(D+1/2)}
\langle \Phi_4(\infty)\Phi_3(1)\Phi_1(q)\Phi_2 (0)\rangle^L_{\tilde{\Delta}^{NS}}=
\frac{q^{\frac{1}{2}(1+\eta-\eta^{(2)}-\eta^{(1)})}}{ \eta}
\,_2F_1\left(A+\frac{1}{2},B+\frac{1}{2};\eta+1;q\right)
\eea
These results are
in agreement with \cite{Belavin:2006zr}. \\
By applying (\ref{Z(0,1)(0,0)(0,0)}) for (\ref{AGTbosfer++})
we get the conformal blocks with two $R$ fields and two $NS$ fields
\bea
\label{NSRRNS(DR)}
&\langle R^{+}_2(\infty) \Phi_1(1) \Phi_4(q) R^{+}_3(0) \rangle^L_{\Delta^R}=
\nonumber \\
&q^{\frac{1}{2}(\eta-\eta^{(3)}-\eta^{(4)})}(1-q)^{-\frac{1}{2} (\eta ^{(1)}-\eta^{(2)}-\eta ^{(3)}+\eta ^{(4)}-1)}\, _2F_1\left(A,B;\eta+\frac{1}{2};q\right)
\eea
The intermediate  field is a Ramond field.

As it was already mentioned
 the conformal blocks with four $R$ fields are expressed in terms of $H_{\pm}$, $\tilde{H}_{\pm}$, $F_{\pm}$, $\tilde{F}_{\pm}$.
 Their connection  to the instanton partition is given in (\ref{Zfor_equal})-(\ref{Zfor_diff1}). Applying (\ref{Z(0,0)(0,1)(0,1)})-(\ref{Z(1,1)(1,0)(0,1)}), we can derive them (see appendix (\ref{H12})-(\ref{F22})).
  Their  expressions get slightly simplified when one takes  $q=\sin^2 (t)$ with $t\in\left( 0\,,\frac{\pi }{2}\right)$.
\bea
\label{RRRRF}
&H_{-}^L(\sin ^2(t))=\frac{\epsilon _2}{\epsilon _1}\frac{ \cos \left(\frac{t}{2}\right) \left(E\left(\sin ^2(t)\right)-\cos (t) K\left(\sin ^2(t)\right)\right) \, _2F_1\left(A+\frac{1}{2},B+\frac{1}{2};\eta +1;\sin ^2(t)\right)}{\pi  \eta   \sqrt[4]{\cos (t)}}
\\
&\tilde{H}_{-}^L(\sin ^2(t))=-\frac{\epsilon _2}{\epsilon _1}\frac{ \sin (t) \left(\cos (t) K\left(\sin ^2(t)\right)+
E\left(\sin ^2(t)\right)\right) \, _2F_1\left(A+\frac{1}{2},B+\frac{1}{2};\eta +1;\sin ^2(t)\right)}{\sqrt{2} \pi  \eta \sqrt[4]{\cos (t)} \sqrt{\cos (t)+1}}\\
&H_{+}^L(\sin ^2(t))=\frac{\sec \left(\frac{t}{2}\right) \left(\cos (t) K\left(\sin ^2(t)\right)+E\left(\sin ^2(t)\right)\right) \, _2F_1\left(A,B;\eta ;\sin ^2(t)\right)}{\pi  \sqrt[4]{\cos (t)}}
\\
&\tilde{H}_{+}^L(\sin ^2(t))=
\frac{\csc \left(\frac{t}{2}\right) \left(\cos (t) K\left(\sin ^2(t)\right)-E\left(\sin ^2(t)\right)\right) \, _2F_1\left(A,B;\eta ;\sin ^2(t)\right)}{\pi  \sqrt[4]{\cos (t)}}
\\
&F_{+}^L(\sin ^2(t))=\frac{\sec \left(\frac{t}{2}\right) \left(\eta  (\cos (t)+1) \, _2F_1\left(A,B+\frac{1}{2};\eta ;\sin ^2(t)\right)-A \sin ^2(t) \, _2F_1\left(A+1,B+\frac{1}{2};\eta +1;\sin ^2(t)\right)\right)}{2 \eta  \sqrt[4]{\cos (t)}}
\\
&F_{-}^L(\sin ^2(t))=\frac{\sec \left(\frac{t}{2}\right) \left(\eta  (\cos (t)+1) \, _2F_1\left(A+\frac{1}{2},B;\eta ;\sin ^2(t)\right)-B \sin ^2(t) \, _2F_1\left(A+\frac{1}{2},B+1;\eta +1;\sin ^2(t)\right)\right)}{2 \eta  \sqrt[4]{\cos (t)}}
\\
&\tilde{F}_{+}^L(-\sin ^2(t))=\frac{\sin (t) \left(A (\cos (t)+1) \, _2F_1\left(A+1,B+\frac{1}{2};\eta +1;\sin ^2(t)\right)-\eta  \, _2F_1\left(A,B+\frac{1}{2};\eta ;\sin ^2(t)\right)\right)}{\sqrt{2} \eta  \sqrt[4]{\cos (t)} \sqrt{\cos (t)+1}}
\\
&\tilde{F}_{-}^L(-\sin ^2(t))=\frac{\sin (t) \left(B (\cos (t)+1) \, _2F_1\left(A+\frac{1}{2},B+1;\eta +1;\sin ^2(t)\right)-\eta  \, _2F_1\left(A+\frac{1}{2},B;\eta ;\sin ^2(t)\right)\right)}{\sqrt{2} \eta  \sqrt[4]{\cos (t)} \sqrt{\cos (t)+1}}
\label{RRRRL}
\eea
\section*{Summary}
With the help of the AGT like correspondence between   $SU(2)$ $\mathcal{N}=2$ super-symmetric gauge theories living on $R^4/Z_2$ space and two dimensional  $\mathcal{N}=1$ SLFT proposed in \cite{Belavin:2011tb,Belavin:2012aa}, analytic expressions are found for the various four point super-conformal blocks in the light asymptotic limit. Namely we have found light blocks when:
 \begin{itemize}            
 	\item all four insertions are NS fields see (\ref{NSNSNSNS(D)}), (\ref{NSNSNSNS(D+1/2)}) ;
 	\item two of the insertions are NS and the other two are Ramond fields see (\ref{NSRRNS(DR)});
 	\item all four are Ramond fields see (\ref{RRRRF})-(\ref{RRRRL}).
 \end{itemize}
The first result of the list above is not new, it has been found in \cite{Belavin:2006zr} via a direct, 
CFT approach. The remaining cases, to my knowledge, are analyzed for the first time and could be helpful 
for better understanding of the subtleties of the light limit in Ramond sector.   
\section*{Acknowledgments}
I am grateful  to Prof. Rubik Poghossian and Prof. Gor Sarkissian
for introducing me into this field of research, for  helpful discussions and comments. This project was partially supported by Armenian SCS grant 15T-1C058.
\appendix
\section{Restriction rules}
\label{ap0}
 Let us look at (\ref{Zbf}). To see whether a box of $\lambda$($\mu$) is in
 $\lambda^*$($\mu^*$) or not we  replace
\bea
\epsilon_1,\, \epsilon_2\rightarrow 1;\,\, a^{(0)}_i\rightarrow u_i;\,\,
a^{(1)}_i\rightarrow q_i;\,\,a^{(2)}_i\rightarrow v_i\,\,\,\,(i=1,2)
\eea
and evaluate a factor corresponding to a box of $\lambda \,(\mu)$. If the result is   equal to $0\, (mod \, 2)$
then the chosen  box belongs to $\lambda^*$($\mu^*$) otherwise not.
Let as apply  this constraint for each of the bifundamentals appearing in (\ref{F}):
\begin{itemize}
\item For $\quad Z_{bf}(u_i,a_i^{(0)},\varnothing\mid q_j, a_j^{(1)},Y_j)$, a box $s\in Y_j $ is also
 in $Y_j^* $ iff
\bea\label{constr1}
 u_i+q_j+1+L_{\varnothing}(s)+A_{Y_j}(s)=0\,(\text{mod}\,2)\,.
 \eea
 \item For $ Z_{bf}(q_i,a_i^{(1)},Y_i\mid v_j,a_j^{(2)},\varnothing)$ a box $s\in Y_i $ is also in
  $ Y_i^* $ iff
\bea\label{constr2}
 q_i+v_j+1+L_{\varnothing}(s)+A_{Y_i}(s)=0\,(\text{mod}\,2)\,.
 \eea
 \item For $ Z_{bf}(q_i,a_i^{(1)},Y_i\mid q_j\,,a_j^{(1)},Y_j)$\\
  a box $s\in Y_i$ is also in $Y_i^*$ iff
 \bea\label{constr31}
  q_i+q_j+1+L_{Y_j}(s)+A_{Y_i}(s)=0\,(\text{mod}\,2)\,.
\eea
a box $s\in Y_j$ is also in $Y_j^*$ iff
 \bea\label{constr32}
  q_j+q_i+1+L_{Y_i}(s)+A_{Y_j}(s)=0\,(\text{mod}\,2)\,.
\eea
\end{itemize}
where $i,j=1,2$.
\section{Proof of the restrictions on the Young diagrams for
 $^{\lozenge}_LZ^{(0,0)}_{(0,0),(0,0)}$ and $^{\lozenge}_LZ^{(0,0)}_{(0,1),(0,1)}$}
\label{Appendix1}
\begin{figure}[]
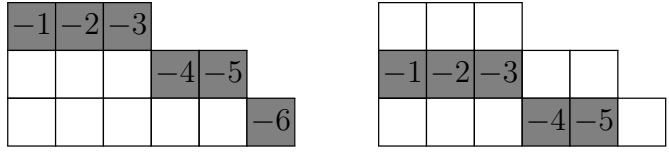

  \centering
\ytableausetup{centertableaux}
\ytableaushort
{\none }
\ytableausetup{nosmalltableaux}
\begin{ytableau}
*(gray)-1  & *(gray)-2& *(gray)-3  \\
*(white) & *(white) &*(white)&*(gray)-4&*(gray)-5\\
&*(white)&*(white)&*(white)&*(white)&*(gray)-6
\end{ytableau}
\qquad
\ytableausetup{nosmalltableaux}
\begin{ytableau}
*(white)  & *(white)& *(white)  \\
*(gray) -1& *(gray) -2&*(gray)-3&*(white)&*(white)\\
&*(white)&*(white)&*(gray)-4&*(gray)-5&*(white)
\end{ytableau}
\caption{The left diagram shows that there are $Y_{i,1}$ boxes such that $A_{Y_i}=0$
(painted gray). The numbers are the leg-length of this boxes towards the empty diagram.
 The right diagram shows that there are $Y_{i,2}$ boxes with $A_{Y_i}=1$ (painted grey)
 and again  the numbers are the leg-length of these boxes towards the empty diagram.}
\label{YD2}
\end{figure}

Here we prove, as we mentioned in section \ref{LPF}, that in the light asymptotic limit contribute only diagrams depicted in figure \ref{figYD3}.
We will give all details for the cases of $^{\lozenge}_LZ^{(0,0)}_{(0,0),(0,0)}$ and $^{\lozenge}_LZ^{(0,0)}_{(0,1),(0,1)}$. The proofs for the other cases are quite similar. Let us  compute the factors in (\ref{F}).\\
 Inserting  (\ref{au_light0}) and (\ref{au_light1})
in (\ref{Zbf}), we obtain for  the first factor of the numerator in (\ref{F}):
 \bea
 \label{Zbfeta1}
&Z_{bf}(u_i,a_i^{(0)},\varnothing\mid q_j, a_j^{(1)},Y_j)=\\
&\prod_{s\in Y_j^{*}}\left(\epsilon_1\left((-)^{i+1}\left(\eta_1
-\frac{1}{2}\right)-
\eta_2+(-)^{j}\left(\eta-\frac{1}{2}\right)+
L_{\varnothing}(s)+1\right)+
\epsilon_2\left(-A_{Y_j}(s)+\frac{(-)^{j+1}-(-)^{i+1}}{2}\right)\right)\nonumber
\eea
In the same way the second factor of the numerator in (\ref{F})  is given by
\bea
\label{Zbfeta2}
&Z_{bf}(q_i, a_i^{(1)},Y_i\mid v_j,a_j^{(2)},\varnothing)=\\
&\prod_{s\in Y_i^{*}}\left(\epsilon_1\left((-)^{i+1}\left(\eta-\frac{1}{2}\right)+
(-)^{j}\left(\eta_4-\frac{1}{2}\right)+\eta_3-
L_{\varnothing}(s)-1\right)+
\epsilon_2\left(A_{Y_i}(s)+\frac{(-)^{j+1}-(-)^{i+1}}{2}\right)\right)\nonumber
\eea
and for the denominator of (\ref{F}) we will get
\bea
\label{Zbfeta3}
&Z_{bf}(q_i,a_i^{(1)},Y_i\mid q_j,  a_j^{(1)},Y_j)=\\
&\prod_{s\in Y_i^{*}}\left(\epsilon_1\left(\left((-)^{i+1}-(-)^{j+1}\right)
\left(\eta-\frac{1}{2}\right)
-L_{Y_j}(s)\right)+
\epsilon_2\left(A_{Y_i}(s)+\frac{(-)^{j+1}-(-)^{i+1}}{2}+1\right)\right)\nonumber
\\\nonumber
&\prod_{s\in Y_j^{*}}\left(\epsilon_1\left(\left((-)^{i+1}-(-)^{j+1}\right)
\left(\eta-\frac{1}{2}\right)
+1+L_{Y_i}(s)\right)+
\epsilon_2\left(-A_{Y_j}(s)+\frac{(-)^{j+1}-(-)^{i+1}}{2}\right)\right)\nonumber
\eea
The instanton expansion coefficients (\ref{F}) are proportional to $\epsilon_1^N$.
We will show that $N>0$ for all  pairs of Young diagram, except those depicted in figure \ref{figYD3}. This means that all other diagrams do not contribute in (\ref{instpartition}) in the light limit ($\epsilon_1\to 0$).\\
Note that  in (\ref{Zbfeta1}) for some boxes from $Y_j^{*}$  the coefficient in front of $\epsilon_2$ vanishes.  Denote the number of such boxes by
  $n_1$. Similarly the numbers of boxes of this kind  in (\ref{Zbfeta2}) and (\ref{Zbfeta3}) are denoted by   $n_2$  and  $n_3$ respectively. It is obvious that
 \bea
 \label{N}
 N=n_1+n_2-n_3.
 \eea
First we explain how to compute the number $n_1$. As we mentioned already, (\ref{Zbfeta1}) is proportional to $\epsilon_1$
whenever  the term proportional to $\epsilon_2$ vanishes. This occurs when
\bea
\label{atm-length1}
A_{Y_j}(s)=\frac{1}{2}\left((-)^{j+1}-(-)^{i+1}\right)\,, \qquad s\in Y_j\,.
\eea
Note that the chosen box $s$ belongs to the same diagram towards which its arm-length is evaluated, hence  the arm-length must always be positive or zero.  From (\ref{atm-length1})  we can see that the only
possible values for $i$ and $j$ that give positive or zero arm-lengths  in (\ref{F}) are:
\bea
\label{constraints11}
j=1;\qquad  i=1;\qquad A_{Y_1}(s)=0; \qquad (s\in Y_1)\,,&\\
\label{constraints12}
j=1;\qquad  i=2;\qquad A_{Y_1}(s)=1; \qquad (s\in Y_1)\,,&\\
\label{constraints13}
j=2;\qquad  i=2;\qquad A_{Y_2}(s)=0; \qquad (s\in Y_2)\,.&
\eea
(\ref{constraints11}) implies that only the boxes that have zero arm-length
contribute to $n_1$. It is obvious from the left diagram of figure \ref{YD2}
that there are
 exactly $Y_{1,1}$ boxes in $Y_1$ for
which the arm-length vanishes (here and below we denote by $Y_{i,k}$
the number of boxes in the $k$'th row of diagram $Y_i$). But not all these
boxes obey the restriction  (\ref{constr1}), which can be written as
 \bea
 u_1+q_1+1+L_{\varnothing}(s)=0 \,\,(\text{mod}\,2), \quad A_{Y_1} (s)=0
 \quad (s\in Y_1)\, .
 \eea
\begin{table}
  \centering
 \scalebox{0.65}{
  \begin{tabular}{ | l | c | c |||||c|c|c|||||c|c|r| }
    \hline
     & $Y_{1,1}=2m$ & $Y_{1,1}=2m+1$
     & &$Y_{1,2}=2k$ & $Y_{1,2}=2k+1$
     & &$Y_{2,1}=2l$ & $Y_{2,1}=2l+1$ \\ \hline
    $u_1+q_1=\text{even}$ &$n_{1,1}=m$ & $n_{1,1}=m+1$&
     $u_2+q_1=\text{even}$ &$n_{1,2}=k$ & $n_{1,2}=k$&
     $u_2+q_2=\text{even}$ &$n_{1,3}=l$ & $n_{1,3}=l+1$\\ \hline
    $u_1+q_1=\text{odd}$ &$n_{1,1}=m$  & $n_{1,1}=m$&
    $u_2+q_1=\text{odd}$ &$n_{1,2}=k$  & $n_{1,2}=k+1$ &
    $u_2+q_2=\text{odd}$ &$n_{1,3}=l$ & $n_{1,3}=l$\\ \hline
   $v_1+q_1=\text{even}$ &$n_{2,1}=m$ & $n_{2,1}=m+1$&
     $v_2+q_1=\text{even}$ &$n_{2,2}=k$ & $n_{2,2}=k$&
     $v_2+q_2=\text{even}$ &$n_{2,3}=l$ & $n_{2,3}=l+1$\\ \hline
    $v_1+q_1=\text{odd}$ &$n_{2,1}=m$  & $n_{2,1}=m$&
    $v_2+q_1=\text{odd}$ &$n_{2,2}=k$  & $n_{2,2}=k+1$ &
    $v_2+q_2=\text{odd}$ &$n_{2,3}=l$ & $n_{2,3}=l$\\ \hline
  \end{tabular}}
  \caption{Depending on $q_i$, $u_i$ and $v_i$, $n_1$ and $n_2$
  take different values. One can get them form this table by
  $n_1=n_{1,1}+n_{1,2}+n_{1,3}$ and $n_2=n_{2,1}+n_{2,2}+n_{2,3}$.}
     \label{tablenum}
\end{table}
From the first picture of figure \ref{YD2} one can see that
$L_{\varnothing}(s)=-1,-2,\dots ,-Y_{1,1}$. Using this we obtain the number of boxes in $Y_{1,1}$ which are in $Y_1^{*}$, denoted by $n_{1,1}$. The results are presented in table \ref{tablenum}.
Correspondingly, the number of boxes satisfying (\ref{constraints12})  with unit arm-lengths in $Y_1$ is equal to $Y_{1,2}$, and  finally,
 the number of the boxes obeying (\ref{constraints13})  with zero arm-lengths in $Y_2$ is equal to $Y_{2,1}$.
But not all of $Y_{1,2}$ and $Y_{2,1}$ boxes are
in $Y_1^{*}$ and $Y_2^{*}$ respectively. We should impose also the constraint (\ref{constr1}). With the same steps  one can
get the number of boxes in  $Y_1^{*}$ and $Y_2^{*}$ denoted by $n_{1,2}$ and $n_{1,3} $ correspondingly. The results  again are summarized in table
\ref{tablenum}. Obviously
\bea
n_1=n_{1,1}+n_{1,2}+n_{1,3}\,.
\eea

 Now let us compute $n_2$. From (\ref{Zbfeta2}) we see that the term proportional to
$\epsilon_2$ vanishes if
\bea
\label{atm-length2}
A_{Y_i}(s)=\frac{1}{2}\left((-)^{i+1}-(-)^{j+1}\right) \qquad s\in Y_i
\eea
where again  the arm-length is towards its own diagram. This means that it is always positive or zero. Therefore
 \bea
\label{constraints21}
&&i=1;\qquad  j=1;\qquad A_{Y_1}(s)=0; \qquad (s\in Y_1)\,;\\
\label{constraints22}
&&i=1;\qquad  j=2;\qquad A_{Y_1}(s)=1; \qquad (s\in Y_1)\,;\\
\label{constraints23}
&&i=2;\qquad  j=2;\qquad A_{Y_2}(s)=0; \qquad (s\in Y_2)\,.
\eea
Again in the $Y_1$
diagram  there are $Y_{1,1}$ and $Y_{1,2}$ boxes  with zero and unit arm-length  and $Y_{2,1}$ boxes in $Y_2$ with zero   arm-length  (see figure \ref{YD2}).
All the boxes that contribute to $n_2$ must obey
(\ref{constr2}). The results is displayed in table \ref{tablenum}.

Let us calculate $n_3$. In (\ref{Zbfeta3})
the therm proportional to $\epsilon_2$ vanishes if
\bea
\label{atm-length3}
&&A_{Y_i}(s)=\frac{1}{2}\left((-)^{i+1}-(-)^{j+1}\right)-1\,; \qquad (s\in Y_i)\,,\\
&&A_{Y_j}(s)=\frac{1}{2}\left((-)^{j+1}-(-)^{i+1}\right)\,;\, \,\,\quad\qquad (s\in Y_j)\,,
\eea
Again both arm-lengths should be positive. This implies
 \bea
 \label{constraints30}
i=1;\qquad  j=2;\qquad A_{Y_1}(s)=0; \qquad (s\in Y_1),&\\
\label{constraints31}
j=1;\qquad  i=1;\qquad A_{Y_1}(s)=0; \qquad (s\in Y_1),&\\
\label{constraints32}
j=1;\qquad  i=2;\qquad A_{Y_1}(s)=1; \qquad (s\in Y_1),&\\
\label{constraints33}
j=2;\qquad  i=2;\qquad A_{Y_2}(s)=0; \qquad (s\in Y_2),&
\eea
Let us apply the  constraint (\ref{constr31}) and (\ref{constr32})  for the boxes defined above. The result
is
\bea
\label{dencon1}
&&\text{$s\in Y_1$ with $A_{Y_1}(s)=0$ is also in $Y_1^*$ if } q_2+q_1+1+L_{Y_2}(s)=0\,(\text{mod 2})\, ;\\
\label{dencon2}
&&\text{$s\in Y_1$ with $A_{Y_1}(s)=0$ is also in $Y_1^*$ if } 1+L_{Y_1}(s)=0\,(\text{mod 2})\, ;\\
\label{dencon3}
&&\text{$s\in Y_1$ with $A_{Y_1}(s)=1$ is also in $Y_1^*$ if } q_2+q_1+L_{Y_2}(s)=0\,(\text{mod 2}) \, ;\\
\label{dencon4}
&&\text{$s\in Y_2$ with $A_{Y_2}(s)=0$ is also in $Y_2^*$ if } 1+L_{Y_2}(s)=0\,(\text{mod 2})\, ,
\eea
Let us denote by $n_{3,j}$ $j=1,2,3,4$ the number of boxes that obey
(\ref{dencon1})-(\ref{dencon4}) correspondingly. Obviously
 \bea
 n_3=n_{3,1}+n_{3,2}+n_{3,3}+n_{3,4}\,.
 \eea
 It is not difficult to see from (\ref{dencon1})-(\ref{dencon4}) that  $n_{3,j}$ obey the constraints
\bea
\label{dconstr1}
&&\text{For both } Y_{1,1}=2m \text{ or } Y_{1,1}=2m+1\,, \quad n_{3,2}\leq m\,;\\
\label{dconstr3}
&&\text{For both } Y_{2,1}=2l \text{ or } Y_{2,1}=2l+1\,, \quad n_{3,4}\leq l\,;\\
\label{dconstr2}
 && n_{3,1}+n_{3,2}\leq Y_{1,1}\,.
\eea
The first two constraints are a consequence of (\ref{dencon2}) and (\ref{dencon4}) respectively. The third constraint can be seen from (\ref{dencon1}) and (\ref{dencon2}).  \\
{\bf The case $^{\lozenge}F^{(0,0)}_{(0,0),(0,0)}$}.\\
From the  above analysis it is  obvious that $N$ depends on the parity (odd or even) of the numbers
$Y_{1,1}$, $Y_{1,2}$ and $Y_{2,1}$.
We will consider each case separately.
\begin{enumerate}
\item If $Y_{1,1}=2m$, $Y_{1,2}=2k$, $Y_{2,1}=2l$. Using table \ref{tablenum}
for $n_1$ and $n_2$ and (\ref{dconstr3}), (\ref{dconstr2}) for $n_3$ we will get
\bea
n_1+n_2=2m+2k+2l,\quad \text{and}\quad n_3\leq 2m+2k+l\,.
\eea
Substituting  this into (\ref{N}) we obtain  $N\ge l$.
In the light asymptotic limit $\epsilon_1\to 0$  the contribution of  a pair of
 diagrams for which $N>0$ is negligible compared to the case with $N=0$. Thus we are interested in pairs of diagrams for which $l=0$.
 This means that $Y_{2,1}=0$. Recalling  that  $Y_{2,1}$ is the number of boxes in the first row of $Y_2$,  we obtain that $Y_{2}$ is an empty Young diagram.\\
  Using (\ref{dencon3}) we can express $n_{3,3}$ in terms of  $Y_{1,2}$
and get $n_3\leq2m
+k$ thus,  $N\ge k$ and  $k=0$, $Y_{1,2}=0$, hence $Y_{2}$
 is a one row diagram with $2m$ boxes.
\item    $Y_{1,1}=2m$, $Y_{1,2}=2k$, $Y_{2,1}=2l+1$
 \bea
 n_1+n_2=2m+2k+2l+2\qquad \text{and} \qquad n_3 \leq2m+2k+l
 \eea
 so that $N\ge l+2$ and thus $N>0$. The contribution of these pairs
 in the instanton partition function (\ref{instpartition}) is negligible compared
to the first case where we had pairs of diagrams with $N=0$.
\item  If  $Y_{1,1}=2m$, $Y_{1,2}=2k+1$, $Y_{2,1}=2l+1$ then
 \bea
 n_1+n_2=2m+2k+2l+2\qquad \text{and} \qquad n_3 \leq2m+2k+1+l
 \eea
 so, $N>0$ and in this case there is no contribution.

\item  $Y_{1,1}=2m$, $Y_{1,2}=2k+1$, $Y_{2,1}=2l$ then
 \bea
 n_1+n_2=2m+2l+2k\qquad \text{and} \qquad n_3 \leq2m+2k+1+l
 \eea
 so we have two possibilities $l=0,1$ that may give a non positive $N$.
\begin{enumerate}
\item When $l=0$ $Y_2$ is empty, then by using (\ref{dencon3})
 \bea
 n_1+n_2=2m+2k\qquad \text{and} \qquad n_3 \leq 2m+k\,.
 \eea
 It seems that for  $k=0$, which is  $Y_{1,2}=1$, one may have a contribution in the partition function. For this case we are able to calculate $n_3$ precisely
 using (\ref{dencon1})-(\ref{dencon4}). The result is $n_3=2m-1$. This means that in fact  $N=1$, thus  we get no contribution.
\item When $l=1$, a careful examination shows that
 $N>0$, therefore no contribution too.
\end{enumerate}
 \item  $Y_{1,1}=2m+1$, $Y_{1,2}=2k$, $Y_{2,1}=2l$ then
 \bea
 n_1+n_2=2m+2+2k+2l\qquad \text{and} \qquad n_3 \leq 2m+1+2k+l
 \eea
 so $N>0$, no contribution.
\item If $Y_{1,1}=2m+1$, $Y_{1,2}=2k$, $Y_{2,1}=2l+1$ then
\bea
n_1+n_2=2m+2+2k+2l+2\qquad \text{and} \qquad n_3 \leq 2m+1+2k+l
\eea
so $N>0$, no contribution.
\item  If $Y_{1,1}=2m+1$, $Y_{1,2}=2k+1$, $Y_{2,1}=2l$ then
\bea
n_1+n_2=2m+2+2k+2l\qquad \text{and} \qquad n_3 \leq 2m+1+2k+1+l
\eea
Thus the only possibility is $l=0$.
This means that $Y_{2,1}=0$ so $Y_{2}$ is an empty Young diagram.\\
Using (\ref{dencon3}) we can see that $n_3 \leq 2m+1+k$
which means that $N>0$, thus no contribution.
\item  If $Y_{1,1}=2m+1$, $Y_{1,2}=2k+1$, $Y_{2,1}=2l+1$ then
\bea
n_1+n_2=2m+2+2k+2l+2\qquad \text{and} \qquad n_3 \leq 2m+1+2k+1+l\qquad
\eea
so $N>0$, no contribution.
 \end{enumerate}
We conclude  that $Y_2$ is empty and $Y_1$ is a one row diagram with even number of boxes.\\
{\bf The case $^{\lozenge}F^{(0,0)}_{(0,1),(0,1)}$}.
\begin{enumerate}
\item If $Y_{1,1}=2m$, $Y_{1,2}=2k$, $Y_{2,1}=2l$. Using table \ref{tablenum}
for $n_1$ and $n_2$ and (\ref{dconstr3}), (\ref{dconstr2}) for $n_3$ we will get
\bea
n_1+n_2=2m+2l+2k,\quad \text{and}\quad n_3\leq 2m+2k+l\,,
\eea
$l=0$ and $Y_2$ is an empty diagram.
By using (\ref{dencon3}) we can express $n_{3,3}$ in terms of  $Y_{1,2}$,
thus we get $n_3\leq 2m+k$ , hence  $k=0$, $Y_{1,2}=0$ so, $Y_{1}$
 is a one row diagram with $2m$ boxes.
\item    $Y_{1,1}=2m$, $Y_{1,2}=2k$, $Y_{2,1}=2l+1$
 \bea
 n_1+n_2=2m+2k+2l\qquad \text{and} \qquad n_3 \leq2m+2k+l\,,
 \eea
 thus, $l=0$, $Y_{2,1}$ may be possible. Using (\ref{dencon3}) we get that 
 $n_3 \leq2m+k$, hence $k=0$. One can check that when $Y_1$ has one row with even number 
 of boxes and $Y_2$ one column with even number of boxes then $N=0$. So this kind of pairs do contribute.
\item  If $Y_{1,1}=2m$, $Y_{1,2}=2k+1$, $Y_{2,1}=2l+1$ then
 \bea
 n_1+n_2=2m+2k+2+2l\qquad \text{and} \qquad n_3 \leq2m+2k+1+l\,,
 \eea
 $N>0$, no contribution.
\item  $Y_{1,1}=2m$, $Y_{1,2}=2k+1$, $Y_{2,1}=2l$ then
 \bea
 n_1+n_2=2m+2k+2+2l\qquad \text{and} \qquad n_3 \leq 2m+2k+1+l\,,
 \eea
 thus, no contribution.
 \item  $Y_{1,1}=2m+1$, $Y_{1,2}=2k$, $Y_{2,1}=2l$, then
 \bea
 n_1+n_2=2m+2+2k+2l\qquad \text{and} \qquad n_3 \leq 2m+1+2k+l
 \eea
 so, no contribution.
\item If $Y_{1,1}=2m+1$, $Y_{1,2}=2k$, $Y_{2,1}=2l+1$ then
 \bea
 n_1+n_2=2m+2+2k+2l\qquad \text{and} \qquad n_3 \leq 2m+1+2k+l
 \eea
 so $N>0$, no contribution.
\item  If $Y_{1,1}=2m+1$, $Y_{1,2}=2k+1$, $Y_{2,1}=2l$ then
 \bea
 n_1+n_2=2m+2+2k+2+2l\qquad \text{and} \qquad n_3 \leq 2m+1+2k+1+l\qquad
 \eea
 $N>0$, thus no contribution.
\item  If $Y_{1,1}=2m+1$, $Y_{1,2}=2k+1$, $Y_{2,1}=2l+1$ then
 \bea
 n_1+n_2=2m+2+2k+2+2l\qquad \text{and} \qquad n_3 \leq 2m+1+2k+1+l\qquad
 \eea
 so $N>0$, no contribution.
 \end{enumerate}
We have shown that  in the light asymptotic limit to the instanton partition function contribute only Young diagrams considered in cases 1 and  2. Combining these two cases we see
that $Y_1$ is a one row diagram with even number of boxes and $Y_2$ is a one column diagram with even number of boxes.

Some instanton partition functions (for example $^{\blacklozenge}Z^{(1,1)}_{(0,0),(0,0)}$) for some set of pairs $Y_1$, $Y_2$ have large expansion coefficients of order $\frac{1}{\e_1}$. These cases are similar to the ones we discussed  above but here
  we should take into account the pairs with $N=-1$ and neglect the ones with
 $N> -1$.
\section{The calculation of $^{\lozenge}_LF^{(0,0)}_{(0,0),(0,0)}$ and $^{\lozenge}_LF^{(0,0)}_{(0,1),(0,1)}$}
\label{Appendix2}
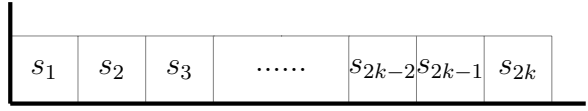
\begin{figure}[]
	\begin{center}
		\begin{tikzpicture}[domain=0:0.5,xscale=0.90,yscale=0.90]
		\draw[style=help lines] (0,0) grid (3,1);
		\draw[style=help lines] (3,1) grid (5,1);
		\draw[style=help lines] (5,0) grid (8,1);
		\node at (0.5,0.5) {$s_1$};
		\node at (1.5,0.5) {$s_2$};
		\node at (2.5,0.5) {$s_3$};
		\node at (4,0.5) {......};
		\node at (5.5,0.5) {$s_{2k-2}$};
		\node at (6.5,0.5) {$s_{2k-1}$};
		\node at (7.5,0.5) {$s_{2k}$};
		\draw[ultra thick,solid ,color=black](0,0)--(0,1.5);
		\draw[ultra thick,solid ,color=black](0,0)--(8.5,0);
		\end{tikzpicture}
	\end{center}
	\caption{The bold line corresponds to  $Y_2$ - an empty diagram;  the thin  lines indicate $Y_1$ - a one row diagram.}
	\label{figc}
\end{figure}
Let as calculate $^{\lozenge}_LF^{(0,0)}_{(0,0),(0,0)}$. As we know  from appendix
 \ref{Appendix1}, $Y_2$ is  empty and $Y_1$ (see figure \ref{figYDSL:subfigure1})
  is a one
 row diagram with even number of boxes.
Let us look at \\ $Z_{bf}(a_2^{(0)},\varnothing\mid a_1^{(1)},Y_1)$. By using
(\ref{Zbfeta1})
we will get
\bea
\label{ex1zbf1}
 Z_{bf}(a_2^{(0)},\varnothing\mid a_1^{(1)},Y_1)=\prod_{s\in Y_2^*}
\left(\epsilon_1\left(-\eta_1-\eta_2-\eta+2+L_{\varnothing}(s)\right)
+\epsilon_2\right)\,,
 \eea
 where we used the fact that the arm-length  $A_{Y_1}(s)=0$ when $s\in Y_1^*$.
  One can see from figure \ref{figc}  that  $L_{\varnothing}(s_1)=-1$,
  $L_{\varnothing}(s_2)=-2$ ... $L_{\varnothing}(s_{2k})=-2k$.
If a box of  $Y_1$ is also in $Y_1^*$ we must use
(\ref{constr1}) which, in this case can be written as $1+L_{\varnothing}(s_{j})=0\,(\text{mod}\,2)$. We see
 that the leg-lengths must be  odd numbers so, $Y_1^*=\{s_1,s_3,\dots s_{2j-1},
 \dots,s_{2k-1}\}$. Thus $L_{\varnothing}(s_{2j-1})=1-2j$ where $j=1, \dots ,k$.
  Inserting this into (\ref{ex1zbf1}) we will get
\bea
Z_{bf}(a_2^{(0)},\varnothing\mid a_1^{(1)},Y_1)=\prod_{j=1}^{k}
\left(\epsilon_1\left(-\eta_1-\eta_2-\eta+3-2j\right)
+\epsilon_2\right)\,.
\eea
The next step is to take $\epsilon_1\to 0$. The result is
\bea
Z_{bf}(a_2^{(0)},\varnothing\mid a_1^{(1)},Y_1)\xrightarrow{\small{ \epsilon_1\to 0}}\epsilon_2^k\,, 
\label{ex.e0.1t}
\eea
all the other bifundamentals are derived with the same steps. Here are the results:
\bea
&&Z_{bf}(a_1^{(0)},\varnothing\mid a_1^{(1)},Y_1)\xrightarrow{\small{ \epsilon_1\to 0}}\epsilon_1^k
\prod_{j=1}^{k}
\left(\eta_1-\eta_2-\eta+2-2j\right)\,;\\
&&Z_{bf}(a_1^{(1)},Y_1\mid a_2^{(2)},\varnothing)\xrightarrow{\small{ \epsilon_1\to 0}}
(-\epsilon_2)^k\,;\\
&&Z_{bf}(a_1^{(1)},Y_1\mid a_1^{(2)},\varnothing)\xrightarrow{\small{ \epsilon_1\to 0}}\epsilon_1^k
\prod_{j=1}^{k}
\left(-\eta_4+\eta_3+\eta-2+2j\right)\,.
\label{ex.e0.2t}
\eea
To get the light asymptotic limit for the
 denominator of (\ref{instpartition}) one must use (\ref{Zbfeta3}) and the constraint rules 
(\ref{constr31}) and (\ref{constr32}). The result will be
\bea
\label{ex.e0.3t}
&&Z_{bf}(a_2^{(1)},\varnothing\mid a_1^{(1)},Y_1)\xrightarrow{\small{ \epsilon_1\to 0}}\epsilon_2^k\,;\\
&&Z_{bf}(a_1^{(1)},Y_1\mid a_2^{(1)},\varnothing)\xrightarrow{\small{ \epsilon_1\to 0}}\epsilon_1^k
 \prod_{j=1}^{k}
\left(2 \eta-2+2j\right)\,;\\
&&Z_{bf}(a_1^{(1)},Y_1\mid a_1^{(1)},Y_1)\xrightarrow{\small{ \epsilon_1\to 0}}
(\epsilon_2\epsilon_1)^k \prod_{j=0}^{k-1}\left(2+2j\right)\,.
\label{ex.e0.4t}
\eea
Now taking the product of (\ref{ex.e0.1t})-(\ref{ex.e0.2t}) and dividing
it to the product of (\ref{ex.e0.3t})-(\ref{ex.e0.4t})
one  gets (\ref{F(0,0)(0,0)(0,0)}).

Now I will derive $F^{(0,0)}_{(0,1),(0,1)}$. As we know  from appendix \ref{Appendix1}, $Y_2$
 is a Young diagram with only one column (see \ref{figYDSL:subfigure2})
containing $2m$ boxes and $Y_1$ a one row Young diagram (see \ref{figYDSL:subfigure1})
 with $2k$ boxes.
 The bifundamentals  are derived
in the same way as in the first case. The results for the numerator of (\ref{F}) are:
\bea
&&Z_{bf}(a_2^{(0)},\varnothing\mid a_2^{(1)},Y_2)\xrightarrow{\small{ \epsilon_1\to 0}}
(-\epsilon_2)^m \prod_{i=1}^{m}(2i-1)\,;\\
&&Z_{bf}(a_1^{(0)},\varnothing\mid a_2^{(1)},Y_2)\xrightarrow{\small{ \epsilon_1\to 0}}
(-\epsilon_2)^m \prod_{i=1}^{m}(2i-1)\,;\\
&&Z_{bf}(a_2^{(0)},\varnothing\mid a_1^{(1)},Y_1)\xrightarrow{\small{ \epsilon_1\to 0}}
\epsilon_2^k\,;\\
&&Z_{bf}(a_1^{(0)},\varnothing\mid a_1^{(1)},Y_1)\xrightarrow{\small{ \epsilon_1\to 0}}\epsilon_1^k
\prod_{j=1}^{k}\left(\eta_1-\eta_2-\eta+2-2j\right)\,;\\
&&Z_{bf}(a_2^{(1)},Y_2\mid a_2^{(2)},\varnothing)\xrightarrow{\small{ \epsilon_1\to 0}}
\epsilon_2^m \prod_{i=1}^{m}(2i-1)\,;\\
&&Z_{bf}(a_2^{(1)},Y_2\mid a_1^{(2)},\varnothing)\xrightarrow{\small{ \epsilon_1\to 0}}
\epsilon_2^m \prod_{i=1}^{m}(2i-1)\,;\\
&&Z_{bf}(a_1^{(1)},Y_1\mid a_2^{(2)},\varnothing)\xrightarrow{\small{ \epsilon_1\to 0}}
(-\epsilon_2)^k\,;\\
&&Z_{bf}(a_1^{(1)},Y_1\mid a_1^{(2)},\varnothing)\xrightarrow{\small{ \epsilon_1\to 0}}\epsilon_1^k
\prod_{j=1}^{k}
\left(\eta-\eta_4+\eta_3-2+2j\right)
\eea
and for the denominator:
\bea
&&Z_{bf}(a_2^{(1)},Y_2\mid a_1^{(1)},Y_1)\xrightarrow{\small{ \epsilon_1\to 0}}
(-\epsilon_2)^k \epsilon_2^m\prod_{i=1}^{m}2i\,;\\
&&Z_{bf}(a_2^{(1)},Y_2\mid a_2^{(1)},Y_2)\xrightarrow{\small{ \epsilon_1\to 0}}
(-\epsilon_2)^m \epsilon_2^m\prod_{i=1}^{m}2i\prod_{i=1}^{m}(2i-1)\,;\\
&&Z_{bf}(a_1^{(1)},Y_1\mid a_1^{(1)},Y_1)\xrightarrow{\small{ \epsilon_1\to 0}}
(-\epsilon_2)^k \epsilon_1^k\prod_{j=1}^{k}2j\,;\\
&&Z_{bf}(a_1^{(1)},Y_1\mid a_2^{(1)},Y_2)\xrightarrow{\small{ \epsilon_1\to 0}}
(-\epsilon_2)^m \epsilon_1^k\prod_{j=1}^{k}(2\eta-2+2j)\prod_{i=1}^{m}(2i-1)\,,
\eea
by dividing the numerator to the denominator one gets (\ref{F(0,0)(0,1)(0,1)}).
\section{$H_{\pm}$, $\tilde{H}_{\pm}$, $F_{\pm}$, $\tilde{F}_{\pm}$ without variable exchange}
\label{Appendix3}

\bea
\label{H12}
H_{-}^L(q)=\frac{\epsilon _2}{\epsilon _1}
\frac{\sqrt{\sqrt{1-q}+1}  \left(E(q)-\sqrt{1-q} K(q)\right) \, _2F_1\left(A+\frac{1}{2},B+\frac{1}{2};\eta +1;q\right)}{\sqrt{2} \pi  \eta  \sqrt[8]{1-q} }
\eea
\bea
\label{H22}
\tilde{H}_{-}^L(q)=-\frac{\epsilon _2}{\epsilon _1}
\frac{\sqrt{\sqrt{1-q}+1} (1-q)^{3/8} \sqrt{q} \left(\sqrt{1-q} K(q)+E(q)\right)
	\, _2F_1\left(A+\frac{1}{2},B+\frac{1}{2};\eta +1;q\right)}{\sqrt{2} \pi  \eta  \left(-q+\sqrt{1-q}+1\right)}\nonumber\\
\eea
\bea
\label{H11}
H_{+}^L(q)=
\frac{\sqrt{2} \sqrt{\sqrt{1-q}+1} (1-q)^{3/8} \left(\sqrt{1-q} K(q)+E(q)\right) \, _2F_1(A,B;\eta ;q)}{\pi  \left(-q+\sqrt{1-q}+1\right)}
\eea
\bea
\label{H21}
\tilde{H}_{+}^L(q)=
\frac{\sqrt{2} \sqrt{\sqrt{1-q}+1} \left(\sqrt{1-q} K(q)-E(q)\right) \, _2F_1(A,B;\eta ;q)}{\pi  \sqrt[8]{1-q} \sqrt{q}}
\eea
\bea
\label{F11}
F_{+}^L(q)=
\frac{\sqrt{\sqrt{1-q}+1} (1-q)^{3/8}}{\sqrt{2} \eta  \left(-q+\sqrt{1-q}+1\right)}\times\qquad\qquad\qquad\qquad\\
\left(\eta  \left(\sqrt{1-q}+1\right)
\, _2F_1\left(A,B+\frac{1}{2};\eta ;q\right)-A q \, _2F_1\left(A+1,B+
\frac{1}{2};\eta +1;q\right)\right)
\nonumber
\eea
\bea
\label{F12}
F_{-}^L(q)=
\frac{\sqrt{\sqrt{1-q}+1} (1-q)^{3/8} }{\sqrt{2} \eta  \left(-q+\sqrt{1-q}+1\right)}\times\qquad\qquad\qquad\qquad\\
\left(\eta  \left(\sqrt{1-q}+1\right) \, _2F_1\left(A+\frac{1}{2},B;\eta ;q\right)-B q
\, _2F_1\left(A+\frac{1}{2},B+1;\eta +1;q\right)\right)
\nonumber
\eea
\bea
\label{F21}
\tilde{F}_{+}^L(-q)=
\frac{\sqrt{\sqrt{1-q}+1} (1-q)^{3/8} \sqrt{q} }{\sqrt{2} \eta  \left(-q+\sqrt{1-q}+1\right)}\times\qquad\qquad\qquad\qquad\\
\left(A \left(\sqrt{1-q}+1\right) \, _2F_1\left(A+1,B+\frac{1}{2};\eta +1;q\right)-\eta
\, _2F_1\left(A,B+\frac{1}{2};\eta ;q\right)\right)
\nonumber
\eea
\bea
\label{F22}
\tilde{F}_{-}^L(-q)=
\frac{\sqrt{\sqrt{1-q}+1} (1-q)^{3/8} \sqrt{q} }{\sqrt{2} \eta  \left(-q+\sqrt{1-q}+1\right)}\times\qquad\qquad\qquad\qquad\\
\left(B \left(\sqrt{1-q}+1\right) \, _2F_1\left(A+\frac{1}{2},B+1;\eta +1;q\right)-\eta  \, _2F_1
\left(A+\frac{1}{2},B;\eta ;q\right)\right)
\nonumber
\eea
\bibliographystyle{JHEP}
\providecommand{\href}[2]{#2}
\begingroup\raggedright

\endgroup

\end{document}